# A new approach of designing Multi-Agent Systems

## With a practical sample

Sara Maalal
Team of Systems' Architecture, Laboratory of computing, Systems and Renewable Energy
National and High School of Electricity and Mechanic
ENSEM BP 8118, Oasis
Casablanca, Maroc

Malika Addou
Team of Systems' Architecture Laboratory of computing, Systems and Renewable Energy
Hassania School of Public
Works EHTP BP 8108, Oasis
Casablanca, Maroc

*Abstract*—Agent technology is a software paradigm that permits to implement large and complex distributed applications [1]. In order to assist analyzing, conception and development or implementation phases of multi-agent systems, we've tried to present a practical application of a generic and scalable method of a MAS with a component-oriented architecture and agent-based approach that allows MDA to generate source code from a given model. We've designed on AUML the class diagrams as a class meta-model of different agents of a MAS. Then we generated the source code of the models developed using an open source tool called AndroMDA. This agent-based and evolutive approach enhances the modularity and genericity developments and promotes their reusability in future developments. This property distinguishes our design methodology of existing methodologies in that it is constrained by any particular agent-based model while providing a library of generic models [2].

*Keyword- Software agents; Multi-agents Systems (MAS); Analysis; Software design; Modeling; Models; Diagrams; Architecture; Model Driven Architecture (MDA); Agent Unified Modeling Language (AUML); Agent Modeling Language (AML).*

## I. INTRODUCTION

Currently the computer systems are increasingly complex, often distributed over several sites and consist of software interacting with each other or with humans. The need for model human behavior in specific computer programs has prompted officials to use technology that affected the last decade and whose movements are very remarkable. In this context, designing multi-agent systems (MAS) is complex because they require the inclusion of several parts of the system which can often be approached from different angles. We must identify and analyze all system problems to find models for multi-agents to implement and integrate them into a coherent system. This is the software engineering and well justifies the use of a method of analysis, design and development of multi-agents systems [2].

This paper describes a practical example of a new generic model designed for modeling multi-agent systems and based on a class diagram, defining the different types of agents and meeting our needs for development and testing of MAS applications.

## II. MULTI-AGENT SYSTEMS

### A. Definitions

- An agent is a computer system within an environment and with an autonomous behavior made for achieving the objectives that were set during its design [3].

- A multi-agents system is a system that contains a set of agents that interact with communications protocols and are able to act on their environment. Different agents have different spheres of influence, in the sense that they have control (or at least can influence) on different parts of the environment. These spheres of influence may overlap in some cases; the fact that they coincide may cause dependencies reports between agents [4].

The MAS can be used in several application areas such as e-commerce, economic systems, distributed information systems, organizations...

### B. Types of agent

Starting from the definitions cited above, we can identify the following agent types [5]:

- The reactive agent is often described as not being "clever" by itself. It is a very simple component that perceives the environment and is able to act on it. Its capacity meets mode only stimulus-action that can be considered a form of communication.

- The cognitive agent is an agent more or less intelligent, mainly characterized by a symbolic representation of knowledge and mental concepts. It has a partial representation of the environment, explicit goals, it is capable of planning their behavior, remember his past actions, communicate by sending messages, negotiate, etc..

- The intentional agent or BDI (Belief, Desire and Intention) is an intelligent agent that applies the model of human intelligence and human perspective on the world using mental concepts such as knowledge, beliefs, intentions, desires, choices, commitments. Its behavior can be provided by the award of beliefs, desires and intentions.

- The rational agent is an agent that acts in a manner allowing it to get the most success in achieving the tasks they were assigned. To this end, we must have measure of performance, if possible objective associated with a particular task that the agent should run.

- The adaptive agent is an agent that adapts to any changes that the environment can have. He is very intelligent as he is able to change its objectives and its knowledge base when they change.

- The communicative agent is an agent that is used to communicate information to all around him. This information can be made of his own perceptions as it may be transmitted by other agents.

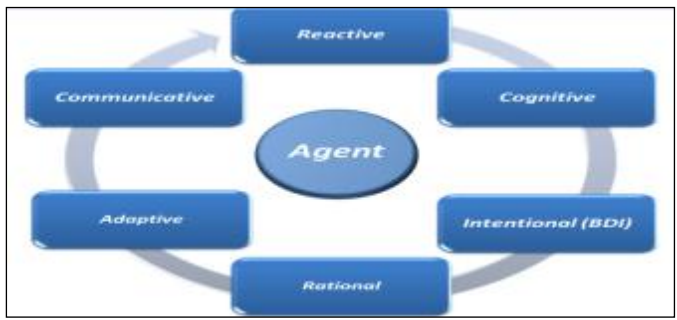

Figure 1. Types of agents

### III. THE DESIGN METHODOLOGIES – STATE OF THE ART

Building high quality software for real-world applications is a difficult task because of the large number and the flexibility of components but also because of the complexity of interconnections required. The role of software engineering is precisely that of providing methodologies that can facilitate control of this complexity. A methodology by definition can facilitate the process of engineering systems. It consists of guides that cover the entire lifecycle of software development. Some are technical guides; others are managing the project [6].

We'll name "method" the approach to use a rigorous process for generating a set of models that describe various aspects of software being developed using a well- defined notation.

To this end, several software engineering paradigms have been proposed, such as object-oriented design patterns, various software architectures. These paradigms fail especially when it concerns the development of complex distributed systems for two reasons: the interactions between the various entities are defined in a too rigid way and there is no mechanism complex enough to represent the organizational structure system [7]. The paradigm of agents and multi-agent systems can be a good answer to these problems, because the agent-oriented approaches significantly increase our ability to model, design and build complex distributed systems [8].

There are many methodologies for analysis and design of multi-agent systems. We cite below some examples of existing methodologies [2]:

- The AAII methodology was developed based on the experience accumulated during the construction of BDI systems. In this methodology, we have a set of templates that, when they have been fully elaborated, define the specifications of agents such as desires, beliefs and intentions [9].

- The first version of Gaia methodology, which modeled agents from the object-oriented point of view, was revisited 3 years later by the same authors in order to represent a MAS as an organized *society* of individuals [10]. In fact, the agent entity, which is a central element of the meta-model of Gaia, can play one or more roles. A role is a specific behavior to be played by an agent (or kind of agents), defined in term of permissions, responsibilities, activities, and interactions with other roles. When playing a role, an agent updates its behavior in terms of services that can be activated according to some specific pre- and post- conditions. In addition, a role is decomposed in several protocols when agents need to communicate some data. The environment abstraction specifies all the entities and resources a multi-agent system may interact with, restricting the interactions by means of the permitted actions [1].

The Gaia methodology gives the possibility to design MAS using an organizational paradigm and to traverse systematically the path that begins by setting out the demands of the problem and to lead to a fairly detailed and immediate implementation [9]. Gaia permits to design a hierarchical non-overlapping structure of agents with a limited depth. From the organizational point of view, agents form teams as they belong to a unique organization, they can explicitly communicate with other agents within the same organization by means of collaborations, and organizations can communicate between them by means of interactions. If inter-organization communication is omitted, coalitions and congregations may also be modeled [1].

However, this methodology is somewhat limited since we can describe MAS with different architectures of agents [9].

- The main contribution of MESSAGE was the definition of meta-models for specification of the elements that can be used to describe each of the aspects that constitute a multi-agent system (MAS) from five viewpoints: organization, agents, goals/tasks, interactions and domain. MESSAGE adopted the Unified Process and centered on analysis and design phases of development [11].

- INGENIAS starts from the results of MESSAGE and provides a notation to guide the development process of a MAS from analysis to implementation [12] [13].

It is both a methodology and a set of tools for development of multi-agent systems (MAS). As a methodology, it tries to integrate results from other proposals and considers the MAS from five complementary viewpoints: organization, agent, tasks/goals, interactions, and environment. It is supported by a set of tools for modeling (graphical editor), documentation and code generation (for different agent platforms). The INGENIAS methodology does not explicitly model social norms, although they are implicit in the organizational viewpoint. Organizational dynamics are not considered i.e., how agents can join or leave the system, how they can form groups dynamically, what their life-cycle is, etc [14]. The authors have developed an agent-oriented software tool called INGENIAS Development Kit (IDK) [15]. It allows to edit consistent models (according to INGENIAS specification) and to generate documented code in different languages such as JADE [16], Robocode, Servlets or Gracias Agents [1].

- Multi-agent systems Software Engineering (MaSE) is a start-to-end methodology that covers from the analysis to the implementation of a MAS [17]. The main goal of MaSE is to guide a designer through the software life-cycle from a documented specification to an implemented agent system, with no dependency of a particular MAS architecture, agent architecture, programming language, or message-passing system.

- AUML (Agent Unified Modeling Language) is an evolving standard for a design methodology to support MAS. It is based on the UML methodology used with object oriented systems. This notation was proposed to adapt the UML's one in order to describe the agent-oriented modeling [18].

AUML provides tools for:

> Specification protocol of interaction between agents,
> Representation of the internal behaviour of an agent,
> Specification of roles, package interface agent, mobility, etc [2].

- The Agent Modeling Language (AML) is a semiformal visual modeling language for specifying, modeling and documenting systems that incorporate concepts drawn from multi-agents systems (MAS) theory [19].

- ASPECS (Agent-oriented Software Process for Engineering Complex Systems) provides a holonic perspective to design MAS [20]. Considering that complex systems typically exhibit a hierarchical configuration, on the contrary to other methodologies, it uses *holons* instead of atomic entities. Holons, which are agents recursively composed by other agents, permit to design systems with different granularities until the requested tasks are manageable by individual entities.

The goal of the proposed meta-model of ASPECS is to gather the advantages of organizational approaches as well as of those of the holonic vision in the modeling of complex system [1].

All these methodologies presented above are still quite recent. They are mainly focused on the analysis phase, whereas design and implementation phases are missing or are redirected to agent-oriented methodologies, which do not offer enough tools to model organizational concepts. Therefore, there is still a gap between analysis and design, which must be specified clearly, correctly and completely [14].

Finally, the maturity of methodologies can be analyzed by the number of systems that have adopted them. Most of analyzed methodologies have associated applications that show their feasibility. These methodologies have been applied in different fields such as medical informatics [21], manufacturing [20] [22], and e-commerce [23]. MaSE and INGENIAS are the most used ones. Unfortunately, the number of real world applications that use agent-oriented methodologies is still low [1].

IV. THE MDA APPROACH

**The MDA (Model Driven Architecture)** proposes a methodological framework and architecture for systems development that focuses first on the functionality and application behavior, without worrying about the technology with which the application will be implemented. The implementation of the application goes through the transformation of business models in specific models to a target platform (Fig.2). One research was done in this area as the dissertation of Jarraya T. [24]

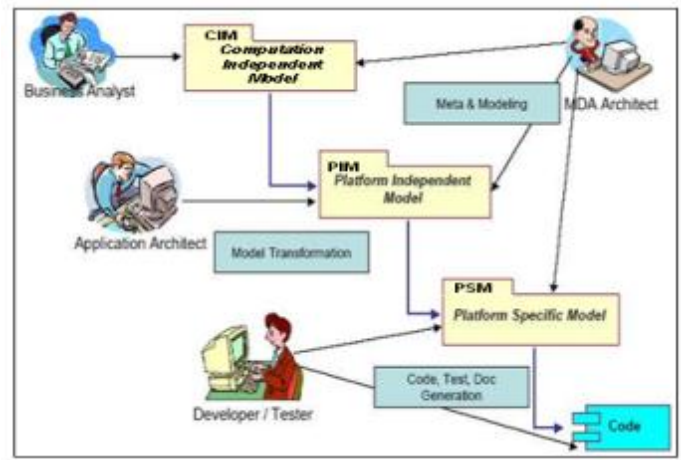

Figure 2. The MDA approach

The business process independent of automation, which comes from the expression of need, is described as a **"CIM" (Computation Independent Model).** The detailed functional analysis, the heart of the process is concentrated in the **"PIM" (Platform Independent Model)**, which, as its name suggests, is strictly independent of the technical architecture and the target language. The **"PSM" (Platform Specific Model)** is the model for engineering design obtained by transformation of PIM by projection on the target technical architecture. It is this model that is based on code generation [5].

The benefits to businesses on the MDA are primarily:

- The fact that architectures based on MDA are ready for technological developments.
- The ease of integrating applications and systems around a shared architecture
- Broader interoperability for not being tied to a platform.

One of the main tools of MDA, we have AndroMDA who takes as its input a business model specified in the Unified Modeling Language (UML) and generates significant portions of the layers needed to build, for example, a Java application [25]. AndroMDA's ability to automatically translate high-level business specifications into production quality code results in significant time savings when implementing Java applications. The diagram below maps various application layers to, for examples, Java technologies supported by AndroMDA [5].

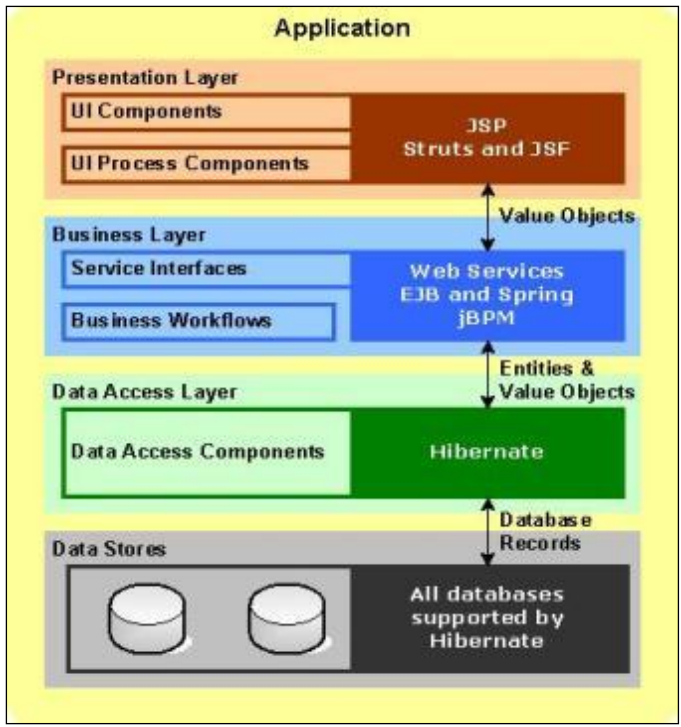

Figure 3. Application layers supported by AndroMDA

- **Presentation Layer:** AndroMDA currently offers two technology options to build web based presentation layers: Struts and JSF. It accepts UML activity diagrams as input to specify page flows and generates Web components that conform to the Struts or JSF frameworks.
- **Business Layer:** The business layer generated by AndroMDA consists primarily of services that are configured using the Spring Framework. These services are implemented manually in AndroMDA-generated blank methods, where business logic can be defined. These generated services can optionally be front-ended with EJBs, in which case the services must be deployed in an EJB container (e.g.,JBoss). Services can also be exposed as Web Services, providing a platform independent way for clients to access their functionality. AndroMDA can even generate business processes and workflows for the jBPM workflow engine (part of the JBoss product line).
- **Data Access Layer:** AndroMDA leverages the popular object-relational mapping tool called Hibernate to generate the data access layer for applications. AndroMDA does this by generating Data Access Objects (DAOs) for entities defined in the UML model. These data access objects use the Hibernate API to convert database records into objects and vice-versa. AndroMDA also supports Enterprise Java Beans EJB3/Seam [26] for data access layer (pre-release).
- **Data Stores:** Since AndroMDA generated applications use Hibernate to access the data, you can use any of the databases supported by Hibernate.

The generation process of AndroMDA is as follows [5] :

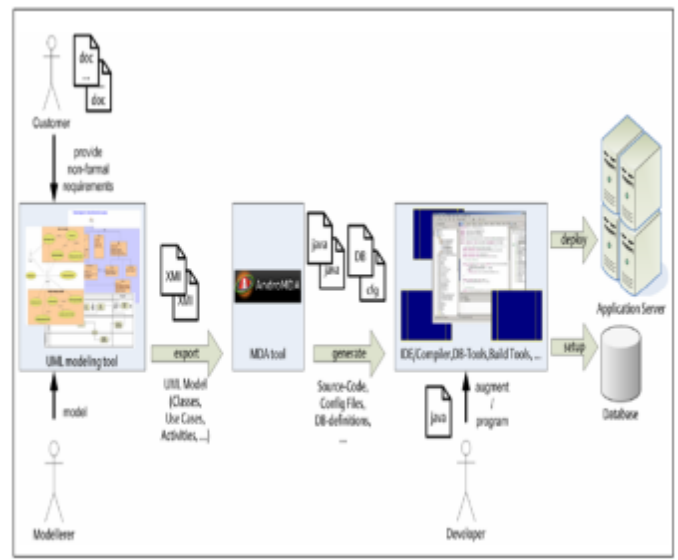

Figure 4. Generation process of AndroMDA

- Preparation of the project in MagicDraw
- Preparing use cases
- Preparation of class diagram
- Preparation of state charts
- Code Generation
- Generating the database
- Deploy the application

V. PROPOSED APPROACH

Our approach is based on model driven architecture (MDA) which aims to establish the link between the existing agent architectures and models or meta-model multi-agent systems that we build based on AUML. Our idea is to offer a design methodology based on agents AUML notation for establishing a generic class diagram that the designer can use to design his system [3]. This diagram is considered as a meta-model which is not generated by any tool and must be defined by the modeler himself.

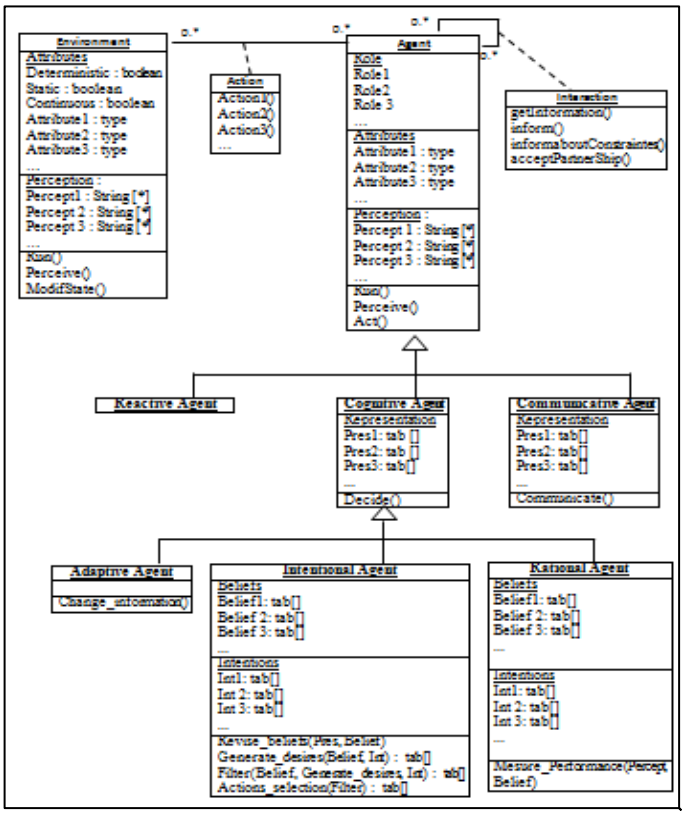

Figure 5. An AUML generic class diagram for a MAS

Our approach has a lot of benefits, it allows:

- Reducing costs and development times for new applications.
- Improving quality of applications.
- Reducing complexity of application development.
- Ability to generate all the necessary components described.
- Modularity and reusability of the developments.
- Coercion by the MDA model.
- Generating a library of generic models.

*A. Description of the AUML generic Class Diagram*

The diagram is conceived in three layers, each one is represented by a relationship between classes: A first part which is a relation between agent and its environment, a second part of specialisation of the agent class, and at the last part, a specialisation of the cognitive agent class [3].

*1- The first part*

The first part consists of two important classes:

- Environment,
- Agent

- **Environment** is an important class on the diagram because it influences all the system. Environment's data is represented by two sections, Attributes and Perceptions. Attributes can be all the information that an environment should have, plus the following common information:

  > *Deterministic* when the next state of the environment is determined in a unique way by the current state and action of the agent, so the environment is deterministic. If the outcome is uncertain (especially if, as a result of action of the agent, the environment can evolve in different ways), we are in the non-deterministic case.

  > *Static* if the environment cannot change its state without the intervention of the agent. The environment is dynamic if its state can change without the action of the agent in the time interval between two perceptions of the agent.

  > *Continuous* if any portion of an environment state to another requires passing through a sequence of intermediate states, otherwise the environment is discrete.

Perception is a section where the designer should determinate all environment perceptions, example: number of agents.

Environment contains several functions allowing to start running, to perceive information from agents linked to it and to modify its state after each action from those agents, that is respectively Run(), Perceive() and ModifState().

- **Agent** is the main class on the diagram that allows the designer to express all agent properties. The constructor of Agents takes three sections: Roles, Attributes and Perception. Roles are agent functionalities. Attributes are all information that an agent should possess. And finally Perception which is a section where the designer should determinate all agents' perceptions about his environment or the other agents.

Agent contains several functions who allows starting running and perceiving information from environment or agents linked to it and to execute all its actions, that is respectively Run(), Perceive() and Act().

The first part consists also of two important association classes:

-Action, between agent and his environment.
-Interaction, between agents.

- **Action** is an association class between agent and environment. It lists all possible actions that an agent can execute on his environment.

- **Interaction** is a reflexive association class between agents. Agent can request information by the getInformation() function and send it by the inform() function. Agent may also deal with some constraints that it is possible to inform by the function informaboutConstraintes(). The acceptance of partnership is added also to the main functionalities of Agent by the function acceptPartnerShip().

2- The second part

The second part represents a specialisation relation of the Agent class. It consists of three important classes:

- Reactive agent,
- Cognitive agent,
- Communicative agent.

- **Reactive agent is** a type of agent. It possesses the same properties of the Agent class.

- **Cognitive agent is** another specialization of the Agent class. In this class, the designer should determinate the representations of the agent that he must have during its execution. The class possesses also one important function "Decide()" where agent can decide to execute an action or not according to his goals.

- **Communicative agent** is the last specialization of the Agent class. Like Cognitive agent class, Communicative agent class has representations but possesses a different function called "Communicate()" where agent must use to communicate his information to the other agents.

3- The third part

The third part represents a specialization relation of the Cognitive agent class. It consists of three important classes:

- Adaptive agent,
- Intentional agent,
- Rational agent.

- **Adaptive agent** is a type of cognitive agent. It possesses the same properties of the Agent class, the knowledge base and the "Decide()" function. As mentioned in the types of agent section above, an adaptive agent is able to change its objectives and its knowledge base as and when these changes. This functionality is expressed by the "Change_information()" function.
- **Intentional agent or BDI Agent is** designed from the "Belief-Desire-Intention" model. It is a type of cognitive agent. In the same case of Adaptive Agent class, this class possesses the same properties of the Agent class, the knowledge base and the "Decide()" function.

In this class, the designer should determinate the agent's beliefs represented by the Beliefs section. The beliefs of an agent are the information that the agent has on the environment and other agents that exist in the same environment. Beliefs may be incorrect, incomplete or uncertain, and because of that, they are different from knowledge of the agent, which is information still true. Beliefs can change over time as the agent by its ability to perceive or interact with other agents, collects more information.

The designer should also determinate the agent's intentions represented by the Intentions section. The intentions of an agent are the actions it has decided to do to accomplish his goals.

To choose the correct agent's beliefs from the incorrect ones, this class offers the "Revise_beliefs(Pres, Belief)" function which is based on the agent's knowledge base and his beliefs. Then, the "Generate_desires(Belief, int)" function comes to generate all the agent's desires that he may be able to accomplish at once. The desires of an agent representing all things the agent would like to see made. An agent may have conflicting desires, in which case he must choose between her desires a subset that is consistent. This subset consists of his desires is identified with the beliefs and the intentions of the agent.

Another function comes after that, the "Filter(Belief, Generate_desires, int)" which filters all those elements above and gives the consistent beliefs, desires and intentions of the intentional agent.

Finally, the agent can select his actions according to this filtration and execute them by the "Actions_selection(Filter)" function.

- **Rational agent** is the last specialisation of the Cognitive Agent class. Like Intentional Agent class, Rational Agent class has the Beliefs and the Intentions sections but possesses just one function called "Mesure_performance(Percept, Belief)" where agent must use to execute his actions as efficient as possible. This function is based both on his perceptions and his beliefs.

*B. The generic UML Class Diagram*

This generic AUML class diagram was subsequently converted into a generic class diagram based on UML notation. This transformation will allow the designer to easily use AndroMDA to generate the source code equivalent to its UML diagram [1].

The passage from AUML to UML was performed by following the steps below:

1. Keep the same titles of classes and associations which constitute the AUML diagram.

2. Assign roles, perceptions, intentions, beliefs and representations of each agent, and any possible additional attributes, in the attributes part of the UML class.

3. Combine all methods or functions in the operations part of the UML class.

We can obtain, in the end, the following result shown in Fig. 6:

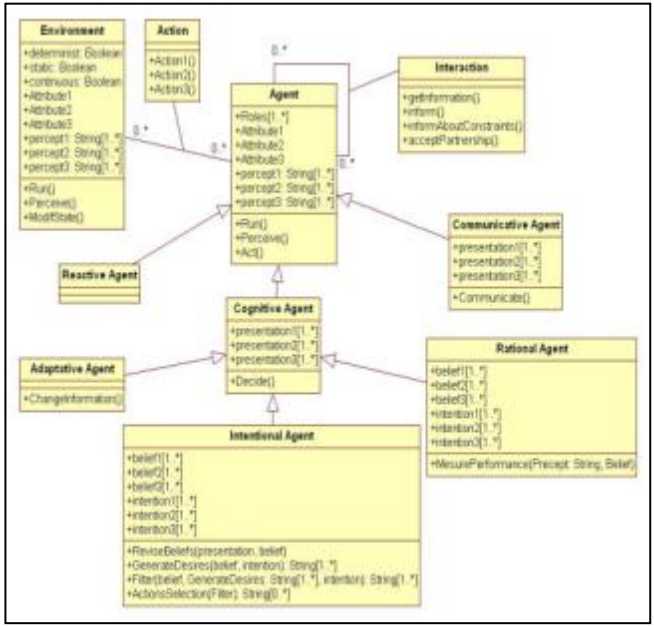

Figure 6. An UML generic class diagram for a MAS

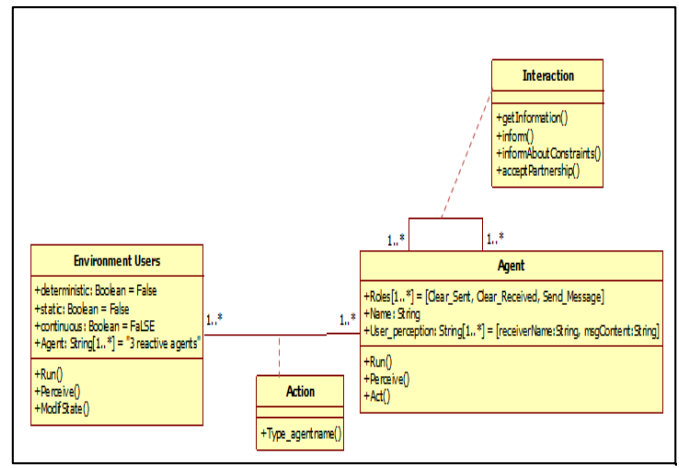

Figure 7. AUML Class diagram for a chat application

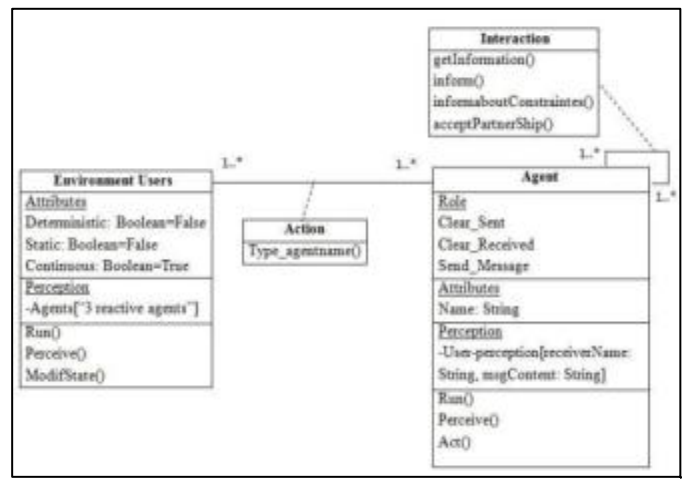

Figure 8. UML Class diagram for a chat application

Our approach can present one desadvantage. It is the complexity of generating a good code source by AndroMDA. The model developed at the design phase, should be reliable in order to build the application and realize its implementation without errors [5].

## V. APPLICATION EXAMPLE

### A. Description

Our proposed AUML class diagram was used for design of one multi-agent system for a Chat Application. This example is designed as follows [5]:

- Three reactive agents: These agents will be the chatters, the interest that these are reactive agents relies on the fact that an agent doesn't react before the declaration of the name of the receiver by the user of the application. Therefore an agent will react to get ready to catch the name and the message and to send it to the appropriate person. He will react also to clear the sent and the received message from their area in his interface.

We can respectively obtain the following AUML and UML diagrams corresponding to this example, shown in the Figures 7 and 8:

### B. Realization

To validate our model for this example, we've tried to download AndroMDA with all the required dependencies (including all profiles referenced by models). Then, we generated our project « ChatAgents » by running « **mvn org.andromda.maven.plugins:andromdaapp-maven plugin:3.4-SNAPSHOT:generate** ». The result of this command is as follows:

When we examine the various folders and files created by the andromdapp plug-in, we will notice files called pom.xml in various folders under ChatAgents. These files make up several Maven projects. In fact, the ChatAgents directory contains a hierarchy of Maven projects as shown below [5].

- **ChatAgents:** This is the master project that controls the overall build process and common properties.
- **mda:** The mda project is the most important sub-project of the application. It houses the ChatAgents UML model under the src/main/uml directory. The mda project is also where AndroMDA is configured to generate the files needed to assemble the application.
- **common:** The common sub-project collects resources and classes that are shared among other sub-projects. These include value objects and embedded values.

classes that use the Spring framework, optionally making use of Hibernate and/or EJBs under the hood. These include entity classes, data access objects, hibernate mapping files, and services.

- **web:** The web sub-project collects those resources and classes that make up the presentation layer.
- **app:** The app sub-project collects those resources and classes that are required to build the .ear bundle.

By opening the file **"ChatAgents.xml"** in MagicDraw, we will be able to build various graphs of our model to generate then the source code of the entire application. Note that AndroMDA can't read MagicDraw 17 models directly. Therefore, you can export it to another file format: EMF-UML2.

After import of AndroMDA profiles to use for our application, we designed our class diagram as shown in Fig.10 as follows [5]:

The result of exporting our **"ChatAgents"** model to EMF-UML2 format is located in the folder C:/ChatAgents/mda/src/main/uml in explorer. Below his content:

- **ChatAgents.xml:** the MagicDraw 17 model file.
- **ChatAgents.uml:** ChatAgents model in EMF/UML2 format. It's the file that will be processed by AndroMDA.
- 10 files ending with .profile.uml: the different profiles used by ChatAgents.uml

Following the definition of our model, the generation of application code is achieved by executing the command "**mvn install**", the result appears as in the figure [5].

Thus, the class **"Chat.java"** is created and can be easily accessed and modified by the developer where he has the ability to implement its operations in the generated code.

We conducted this implementation and got the final result.

Figure 9 : ChatAgents project generation

```
ChatAgents
|
|-- mda
|
|-- common
|
|-- core
|
|-- web
|
+-- app
```

- **core:** The core sub-project collects resources and

## VI. CONCLUSION AND FUTURE SCOPE

The purpose of this paper is to demonstrate the feasibility of our approach to analyze, design and implement multi-agent systems. With AUML modeling and MDA, we can generate all the necessary components described by the class meta-model that we proposed. Which leads us to obtain a generic design based on SOA more or less reusable components using one of the most MDA tools used in development is AndroMDA [27].

In the future, we would like to model another application sample of our model but in a more complex form using cognitive or adaptive agents and in other platforms like C++, Web services, etc. It will help us to validate the efficacy of our proposed approach and lead us to consider it as a generic approach which can be adopted by every type of information system and used for any real world application.

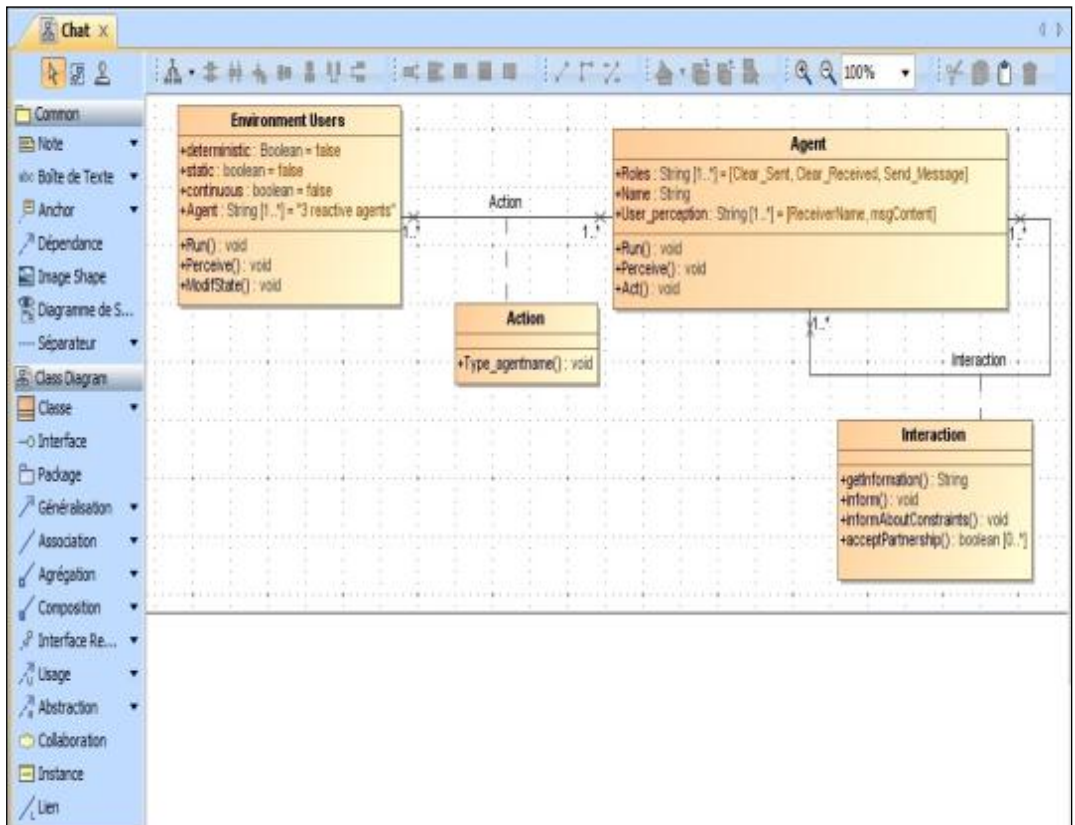

Figure 10 : Class diagram built on MagicDraw 17

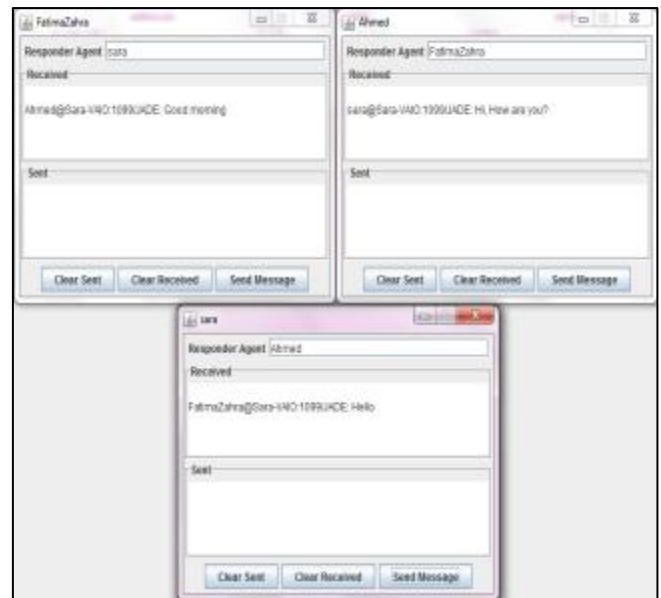

Figure 12. Chat application with three agents

ACKNOWLEDGMENT

I would like to thank to my advisor Ms. M. Addou, Phd. for his invaluable guidance and many useful suggestions during my work on this paper. I would also like to express my gratitude to all those who gave me the possibility to complete this paper.

Figure 11 : Code generation after definition model

## AUTHORS PROFILE

**Sara Maalal** was born in Rabat the Morocco's capital in 1985. She received his professional master in Computer Engineering and Internet (3I), Option: Security Networks and Systems, in 2008 from the Faculty of science of HASSAN II University, Casablanca, Morocco. In 2010 she joined the system architecture team of the National and High School of Electricity and Mechanic (ENSEM: Ecole Nationale Supérieure d'Electricité et de Mécanique), Casablanca, Morocco.
Her actual main research interests concern Designing and modeling Multi-Agent Systems.
Ms. Maalal is actually a Software Engineer in a Moroccan multinational society called Hightech Payment Systems (HPS) which has always proved itself as a leading payment solutions provider.

**Malika Addou** received her Ph.D. in Artificial Intelligence from University of Liege, Liege, Belgium, in 1992. She got her engineer degree in Computer Systems from the Mohammadia School of Engineers (EMI : Ecole Mohammadia des ingénieurs), Rabat, Morocco in 1982. She is Professor of Computer Science at the Hassania School of Public Works (EHTP : Ecole Hassania des Travaux Publics), Casablanca, since 1982.
Her research focuses on Software Engineering (methods and technologies for design and development), on Information Systems (Distributed Systems) and on Artificial Intelligence (especially Multi-Agent Systems technologies).